\documentclass[pre,amsmath,showpacs,twocolumn]{revtex4}
\usepackage{graphicx}
\begin{document}
\title{Agglomeration of Oppositely Charged Particles in Nonpolar Liquids}
\author{Jochen H. Werth$^{1,2}$, Henning Knudsen$^{1}$, and Haye Hinrichsen$^{2}$}
\affiliation{$^1$ Theoretische Physik, Fachbereich 10,
     Universit{\"a}t Duisburg-Essen,
     47048 Duisburg, Germany}
\affiliation{$^2$ Fakult\"at f\"ur Physik und Astronomie,
     Universit{\"a}t W\"urzburg, Am Hubland,
     97074 W\"urzburg, Germany}

\date{January 5, 2004}

\begin{abstract}
We study the aggregation of insulating electrically charged spheres suspended in a nonpolar liquid. Regarding the van der Waals interaction as an irreversible sticking force, we are especially interested in the charge distribution after aggregation. Solving the special case of two oppositely charged particles exactly, it is shown that the surface charges either recombine or form a residual dipole, depending on the initial condition. The theoretical findings are compared with numerical results from Monte Carlo simulations.
\end{abstract}
\parskip 2mm


\maketitle
\def\makevector#1{\vec{#1}}
\def\rvec{\makevector{r}}
\def\Fvec{\makevector{F}}
\def\xivec{\makevector{\xi}}
\def\jvec{\makevector{j}}
\def\vvec{\makevector{v}}
\def\uvec{\makevector{u}}
\def\evec{\makevector{e}}


\section{Introduction}
Fine powders with particles on the micrometer scale play an increasing role in diverse technological applications, ranging from solvent-free varnishing to inhalable drugs~\cite{Review1}. A major problem in this context is the tendency of the particles to clump due to mutual van der Waals forces~\cite{Sontag}, leading to the formation of aggregates. In many applications, however, the aggregates should be sufficiently small with a well-defined size distribution. 
A promising approach to avoid clumping is to coat the powder by nanoparticles. The small particles act as spacers between the grains, reducing the mutual van-der-Waals forces and thereby increasing the flowability of the powder. However, the fabrication of coated powders is a technically challenging task since the nanoparticles themselves have an even stronger tendency to clump, forming large aggregates before they are deposited on the surface of the grain. One possibility to delay or even prevent aggregation is the controlled use of electrostatic forces. As shown in Ref.~\cite{Wirth01} this can be done by charging the nanoparticles and the grains oppositely. On the one hand the repulsive interaction between equally charged nanoparticles suppresses further aggregation once the Coulomb barrier between the flakes has reached the thermal
energy~\cite{Werth03}. On the other hand, attractive forces between the nanoparticles and the grains support the coating process.

\begin{figure}
\includegraphics[width=70mm]{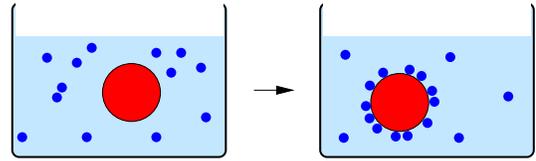}
\caption{
\label{FIG1} \footnotesize
Schematic experimental setup of the coating process: Large and small
particles are suspended in a nonpolar liquid. Charging them oppositely
the small particles are preferentially deposited at the surface of the large particles.
} 
\end{figure}

The coating process is most easily carried out if both fractions of particles are suspended in a liquid (see Fig.~\ref{FIG1}). This type of coating processes requires the use of a nonpolar liquid such as liquid nitrogen. In contrast to colloidal suspensions in polar liquids, the charged particles suspended in liquid nitrogen are not screened by electrostatic double-layers. Both the large and the small particles are insulators so that the charges reside on their surface. By choosing different materials and charging them triboelectrically, it is possible to charge the two particle fractions oppositely in a single process\cite{Wirth01}.
\begin{figure}
\includegraphics[width=54mm]{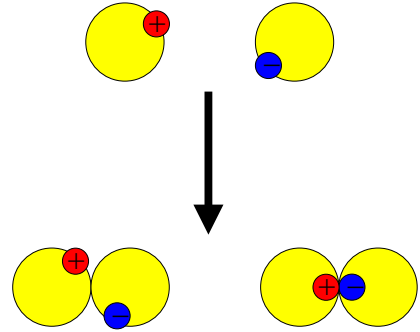}
\caption{
\label{FIG2} \footnotesize
If two oppositely charged particles aggregate due to van der Waals forces, the charges may either recombine or form a residual dipole.
} 
\end{figure}      

\begin{figure}
\includegraphics[height=40mm]{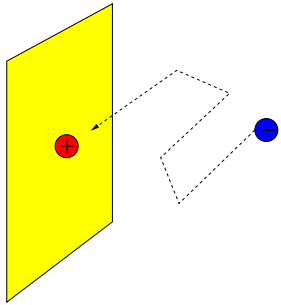}
\caption{
\label{FIG3} \footnotesize
Simplified model for electrostatically supported coating:
A negatively charged pointlike particle is attracted by a positive
charge located at the planar surface of an infinitely large
spherical particle.
} 
\end{figure}

What are the morphological properties of the coated surface? Do the
nanoparticles reach their countercharges exactly or do they attach
elsewhere on the surface, forming residual dipoles, as sketched in
Fig.~\ref{FIG2}? In order to address these questions we consider a
simplified situation, where pointlike particles are deposited on the
surface of a plane, representing the surface of an infinitely large
spherical particle (see Fig.\,\ref{FIG3}). One or several positive
charges are located on the planar surface of the big particle,
attracting negatively charged pointlike particles inserted far
away. For simplicity we assume that the particles are inserted one
after another so that mutual interactions during the deposition
process can be neglected. Similarly, we assume that the hydrodynamic
interactions between the particle and the plane can be ignored. Thus
the small particle is subjected to Coulomb forces and Stokes friction
as well as Brownian motion. As shown in Ref.~\cite{Werth03}, the
damping time of suspended nanoparticles is so short that on the time scale relevant for the coating process their motion can be assumed as overdamped, i.e., inertia can be neglected. The van der Waals interaction is interpreted as a purely adhesive force, i.e., once a particle touches the surface of the large particle it sticks 
irreversibly.  

In this study we show that a certain fraction of the particles exactly reach and compensate their countercharges. The remaining particles are distributed around the charges, partly decaying with distance, partly as a constant background. The fraction that exactly reaches the countercharges is determined by the interplay between the magnitude of the charges, the density of the charges, and the diffusion constant. It can be used as a measure to what extent a predefined structure of positive charges at the surface survives during the coating process.

\section{Theoretical Predictions}
\label{TheorySec}

\subsection{Formulation of the problem}

%
%
In order to study the deposition process analytically, one has to solve the equation of motion of the particle subjected to Coulomb forces, Stokes friction, as well as Brownian motion. We work in the overdamped limit which can be justified as follows. On the one hand, the viscous motion of the particle decays exponentially on a time scale 
\begin{equation}
  t_{\text{relax}} = \frac{m}{6 \pi \eta a}\,,
\end{equation}
where $\eta$ is the viscosity of the fluid, $a$ is the particle radius,
and $m$ the particle mass. On the other hand, the typical time scale
for a particle to diffuse thermally by its own diameter is
given by 
\begin{equation}
t_{\text{diff}} = \frac{18 \pi \eta a^3}{k_B T}.
\end{equation}
As shown in Ref.~\cite{Werth03}, under typical experimental conditions in liquid nitrogen
$t_{\text{relax}}$ is always much smaller than $t_{\text{diff}}$, even if
the particles are as small as $1\;$nm. Therefore, on time scales larger
than $t_{\text{diff}}$, the particle performs a random walk guided by
the balance of Coulomb forces and Stokes friction. Such a motion can
be described by the Langevin equation 
\begin{equation}
\frac{\partial}{\partial t} \rvec = \frac{\Fvec_C(\rvec)}{6 \pi \eta
  a} + \xivec(t)\ ,
\label{EQLANGEVIN}
\end{equation}
where $\rvec$ is the position of the particle, $\Fvec_C(\rvec)$ is
the Coulomb force acting on it, and $\xivec(t)$ is a white Gaussian
noise with the correlations 
\begin{equation}
\langle \xi_i(t) \xi_j(t') \rangle = \frac{k_B T}{3 \pi \eta a} \, \delta_{ij} \, \delta(t-t').
\end{equation}
Equivalently, one may formulate the problem in terms of a
Fokker-Planck (FP) equation~\cite{Risken}, which describes the
temporal evolution of the probability distribution $P(\rvec,t)$ to
find the particle at point $\rvec$ and time $t$. The FP equation has
the form 
\begin{equation}
\frac{\partial}{\partial t} P(\rvec,t) = -\vec{\nabla}\cdot \jvec(\rvec,t)\,,
\end{equation}
where
\begin{equation}
\jvec(\rvec,t) = -D\vec{\nabla} P (\rvec,t) + \vvec(\rvec) P (\rvec,t)
\end{equation}
is the probability current, $D$ the diffusion constant, and
\begin{equation}
\vvec(\rvec) = -\frac{q^2 \rvec}{24 \pi^2 \varepsilon \varepsilon_0 \eta a r^3} =
-\frac{Q}{r^2} \, \frac{\rvec}{r}
\end{equation}
is the particle velocity in the overdamped limit.

Rescaling space and time by 
\begin{equation}
\rvec \to \frac{Q}{D} \rvec \,, \qquad
t \to \frac{Q^2}{D^3} t
\end{equation}
and suppressing the arguments 
$\rvec,t$ we obtain the parameter-free dimensionless equation
\begin{equation}
\label{DimlessFP}
\frac{\partial}{\partial t} P =
\nabla^2 P -
\vec{\nabla} P\cdot \uvec-
P \, (\vec{\nabla} \cdot \uvec)
\end{equation}
where $\uvec=-\rvec/r^3$.

\subsection{Solution of the Fokker-Planck equation}
\label{SolFokker}
\begin{figure}
\includegraphics[width=80mm]{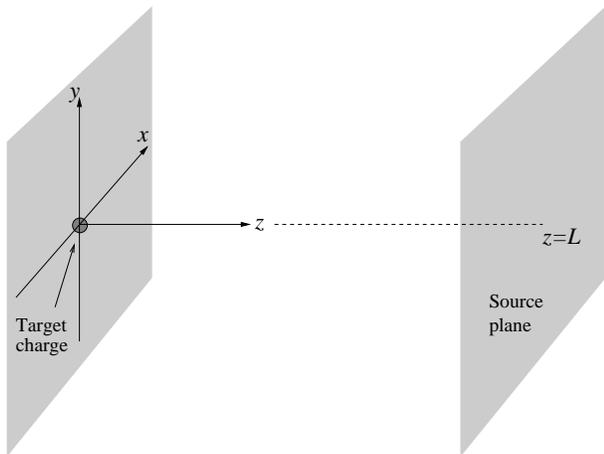}
\caption{
\label{FIG4} \footnotesize
A thermally diffusing particle inserted randomly at finite distance
$z{=}L$ is attracted by the target charge at the origin. Once the
particle touches the target plane $z{=}0$, which represents the large
particle to be coated, it sticks to its surface due to van-der-Waals
forces. The single-particle
problem can be recast as a quasi-stationary flow of a large number of
particles inserted continously along the source plane which are removed
whenever they touch the target plane. 
} 
\end{figure}

In what follows we consider a pointlike particle inserted at a finite
distance $z{=}L$ from the plane with random coordinates $x$ and $y$, 
as shown in Fig.~\ref{FIG4}. The particle diffuses guided by
the Coulomb force until it touches the wall at $z{=}0$, where it sticks  
irreversibly. Our aim is to compute the probability distribution $\rho(r)$
for the particle to touch the wall at a distance $r{=}\sqrt{x^2+y^2}$.
To this end we consider the problem as a \textit{quasi-stationary} process, where
many particles are continuously introduced at the source plane $z{=}L$ and removed
at the target plane $z{=}0$ (see Fig.~\ref{FIG4}). 
Thus the probability distribution $P_s(\rvec)$
to find a particle at position $\rvec$ is a solution of the 
stationary FP equation
\begin{equation}
\label{eq:statfp}
\nabla^2P_s - \vec{\nabla}P_s\cdot\uvec-P_s(\vec{\nabla}\cdot \uvec) = 0
\end{equation}
together with the boundary condition
\begin{equation}
\left. P_s(\rvec) \right|_{z=0} = 0
\end{equation}
and an appropriate source term at $z{=}L$. Taking the limit $L\to\infty$
this problem can be solved exactly in two (2D) and three (3D) spatial
dimensions  (see Appendices \ref{Sol2D} and \ref{Sol3D}). 
In the original variables the stationary probablity distribution is
given by
\begin{equation}
\label{eq:Psol}
P_s(\rvec) = 
\left\{
\begin{array}{ll}
z & \text{in 2D} \\
z(1 + 1/2r) & \text{in 3D}\ .
\end{array}
\right.
\end{equation}
The probabilitiy current in 2D is given by 
\begin{equation}
\jvec  =  -\evec_z-z\rvec/r^3 =
\left(\begin{array}{c}
-xz/r^3 \\ -1-z^2/r^3 \end{array}\right)\ ,
\end{equation}
while in 3D one arrives at a more complicated expression
\begin{equation}
\jvec  =  -\Bigl(1+\frac{1}{2r}\Bigr)\left(\begin{array}{c}
0\\0\\1\end{array}\right)
-\frac{(1+r)}{2}\,\frac{z}{r^4}
\left(\begin{array}{c}
x\\y\\z\end{array}\right)\;.
\end{equation}
Remarkably, in both cases the vector field $\jvec(\rvec)$ exhibits a
separatrix between a region, where the flux lines reach the
target charge at the origin, and another region, where they terminate elsewhere on
the surface (see Fig.~\ref{FIG5}). 
As shown in Appendix~\ref{SolFlux} it is even possible to calculate the
separatrix exactly.
Note that the flux lines and the actual trajectories of
the particles have a different meaning.

\begin{figure}
\includegraphics[width=80mm]{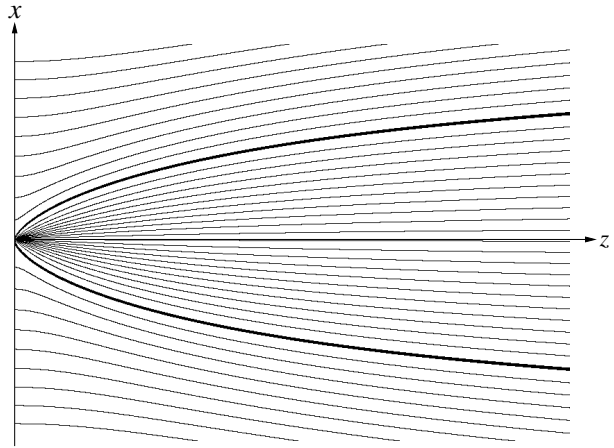}
\caption{
\label{FIG5}
\footnotesize
Flux lines of the vector field $\jvec(\rvec)$ in the $x{-}z$ plane
for the three-dimensional problem. The separatrix
between flux lines terminating at the target charge and those
terminating elsewhere at the wall is represented as a bold line. 
In the two-dimensional case one obtains a qualitatively 
similar flow field.
} 
\end{figure}      
The density of particles $\rho(r)$ reaching the wall at distance $r$
from the target charge is proportional to the normal component of the
flux $\left. j_z(\rvec) \right|_{z=0}$. As shown in Appendix~\ref{SolWall}, we
obtain 
\begin{equation}
\label{DensityResult}
\rho(r) =
\left\{
\begin{array}{ll}
1 + 2\delta(r)  & \text{in 2D} \\
1 + 1/2r + \frac{\pi}{2}\delta(r) & \text{in 3D}\ .
\end{array}
\right.
\end{equation}
Concerning the problem of several target charges, we note that in 2D
the problem is still analytically solvable. For a given charge density
$\lambda(r)$ on the target line, the density of charges reaching the
wall becomes
\begin{equation}
  \rho(r) = 1 + 2\lambda(r)\ .
\end{equation}
This result stems from the fact that in 2D the stationary probability
distribution $P_s(\vec{r}){=}z$, Eq.\,(\ref{eq:Psol}), is independent
of the charge position, or more generally, independent of the
distribution $\lambda$. This can be verified by explicitly inserting
the solution in Eq.\,(\ref{eq:statfp}). The Laplacian $\nabla^2 P_s$ vanishes so that
the resulting equation becomes linear in the Coulomb potential that is
contained in $\vec{u}$. The proposed solution for $P_s$ solves the
equation for any point charge, and due to this linearity, it solves
the equation for any charge distribution. In the physically more
relevant three-dimensional case, however, we resort to numerical
methods to study the effect of more charges.

\section{Numerical  Results}

\subsection{Implementation}

The discretized equation of motion for a single particle can be
derived from the Langevin equation (\ref{EQLANGEVIN}) and is given by
\begin{equation}
d\rvec = -\frac{Q}{r^2}\frac{\rvec}{r}dt + \sqrt{2Ddt}\;{\vec{\cal R}}
\end{equation}
where $Q= Dq^2/(8\pi\varepsilon\varepsilon_0 k_BT)$ is a constant
containing the diffusion constant $D$, dielectric constant
$\varepsilon\varepsilon_0$, particle charge $q$, and thermal energy $k_BT$. $\vec{\cal R}$
is a vector of Gaussian distributed random numbers with unit variance, representing
Brownian motion of the particle. As in the previous section, this equation can be made
dimensionless by rescaling space and time by
$\rvec\rightarrow\frac{Q}{D}\rvec$ and
$t\rightarrow\frac{Q^2}{D^3}t$, leading to 
\begin{equation}
d{\vec r} = -\frac{\rvec}{r^3}dt + \sqrt{2dt}\;{\vec{\cal R}}
\label{EQOFMOTION}
\end{equation} 
As in the FP equation (\ref{DimlessFP}), no free parameter is left in
this equation. Thus, apart from the units of space and time, the
solution is universal.  
The numerical integration of Eq.\,(\ref{EQOFMOTION}) can be
performed easily for a large number of representations of the Brownian
motion $\vec{\cal R}$ and initial positions at the source plane on a workstation. 

\begin{figure}
\includegraphics[width=80mm]{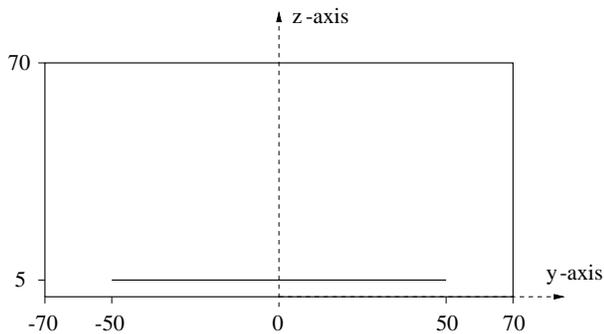}
\caption{
\label{FIG6} \footnotesize
Boundary conditions in the numerical simulation: The picture shows a
two-dimensional cut through the three-dimensional simulation setup. The
$x$-axis points perpendicular to the shown plane towards the reader. 
} 
\end{figure}

A two-dimensional cut through the three-dimensional simulation setup is shown in Fig.\ref{FIG6}. 
One or
several charges are fixed at the $y$-axis. Small particles start
diffusing from the plane ($-50\leq x,y \leq +50, z=+5$) towards the
y-axis. In order to avoid diffusion of particles too far away from the
interesting region, the simulation volume is confined by walls at
$x,y=\pm 70$ and $z=+70$. Particles touching this walls stick
irreversibly. 

\subsection{Comparision of numerical results and theoretical predictions}

First we want to show that in case of a single charge fixed at the
wall our numerical simulations reproduce the exact analytical
solution. This has to be done, since the boundary conditions of our
simulation setup (diffusing particles start from a plane with finite
extent, not far away from the absorbing wall; additional absorbing
walls confine the simulation space) are different from the boundary
conditions of the Fokker-Planck-equation seen above. 

By running the simulation without any attracting charges fixed at the
walls (i.e. particles only diffuse) one can compare simulation results to the homogeneous
distribution of particles on the wall one would expect from solving the FP
equation without any attracting charge. The
numerical results show a homogeneous distribution of particles
hitting the wall in a sufficiently big region. However, approximately 14\% of
the particles hit the additional walls surrounding the simulation
space, leading to a reduced influx of particles to the $z{=}0$
plane. This will be taken into account in the following graphs by
amplification of the incoming particle fluxes to compensate for this
loss of particles. 

\begin{figure}
\includegraphics[width=70mm,angle=-90]{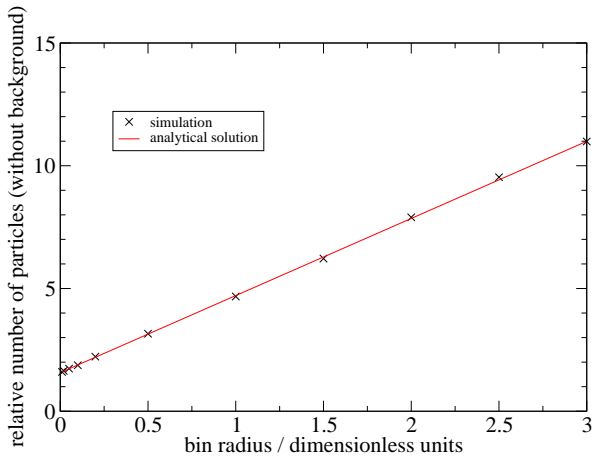}
\caption{
\label{FIG7} \footnotesize
Influx of particles onto the wall within a circular shaped region
around the fixed charge. The straight line represents the analytical
solution: $\frac{\pi}{2}+\pi R$. In both cases, the constant
background is subtracted. 
} 
\end{figure}

In order to check the numerical results for the case of one attracting
charge fixed at the wall, we calculate the influx of particles in a
circular region of radius $R$ around the attracting charge. From
equation (\ref{DensityResult}), the influx is given by
\begin{equation}
\int\limits_0^R \left(1+\frac{1}{2r}\right)2\pi
rdr+\frac{\pi}{2}\delta(r) = \pi R^2 + \pi R + \frac{\pi}{2}.
\end{equation}
Fig.\,\ref{FIG7} shows the influx obtained in the simulation for
different bin sizes $R$ compared to the theoretical influx after
subtraction of the homogeneous background influx. As one can see, both
are in excellent agreement.

\subsection{Numerical results for serveral target charges}

\begin{figure}
\includegraphics[width=80mm]{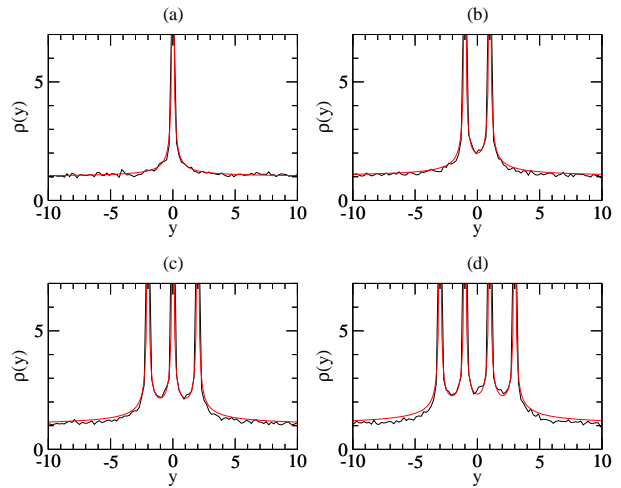}
\caption{
\label{FIG8} \footnotesize
Density profile of particle influx for one (a), two (b), three (c) and
four (d) charges fixed on the y-axis. The bold curves show data from
our simulations, the ligt curves are computed as described in the
text. 
} 
\end{figure}

What changes, if serveral attracting charges are fixed at the wall?
Fig.\,\ref{FIG8} shows simulation results for up to four charges fixed
at the wall. The dimensionless Langevin equation is now given by
\begin{equation}
d{\vec r} = \sum\limits^{n}_{i=1}-\frac{\rvec-\rvec_i}{(r-r_i)^3}dt +
\sqrt{2dt}\;{\vec{\cal R}}\ ,
\label{EQOFMOTION_SERVERAL_TARGETS}
\end{equation} 
where $n$ is the number of charges. 
The charges are always located on the y-axis, separated by
a distance of two dimensionsal units. The boundary conditions are
still the same as shown in Fig.\,\ref{FIG6}. Fig. \ref{FIG8} shows the density
distribution of incoming particles in a small strip around the
y-axis ($|y|\leq0.5$). In each graph the dark line shows simulation
results, while the light curve is given by
\begin{equation}
\rho(x=0,y,z=0)=1+\sum\limits_{i=1}^n\frac{1}{2|y-y_i|},
\end{equation}
where $n$ is the number of charges fixed at the y-axis. As one can
see, the assumed superposition of the $1/r$-shoulders from the
single charge solution fits suprisingly well, since this kind of
superposition is not a solution of the FP-equation. However, one can
see that in the case of three and four charges, the density of
particles near the outer charges is slightly overestimated. 

\begin{figure}
\includegraphics[width=70mm]{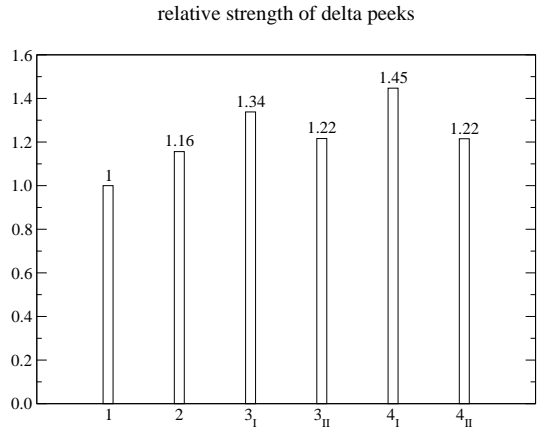}
\caption{
\label{FIG9} \footnotesize
Relative strenght of the delta peaks for one (1), two (2), three
($3_I$ for the inner, $3_{II}$ for the outer peak) and four ($4_I$ for
the inner, $4_{II}$ for the outer peak) charges.  
} 
\end{figure}

The relative strength of $\delta$-peaks for different numbers of
charges is shown in Fig.\,\ref{FIG9}. Again, this data is obtained by
placing a circular bin around each charge location and extrapolating
the strength of the delta peak from a vanishing bin radius. As one can see,
the amplitude of the $\delta$-peak grows by increasing the number of
charges. Also, charges in the center of line always collect more
particles than charges on the edge. 

{}

%

\subsection{Agglomeration of equally sized spherical particles}

So far the limit of one infinitely large particle and one small particle was
examined in detail. Now we address the other limit, namely two spherical
particles of the same size. 

The rotational degree of freedom must be taken into account when the two
particles are of equal size. As for the translational degree of freedom, we
assume that the overdamped limit is valid also for the rotational motion. 
Again, the rotational motion can be described by a Langevin equation, i.e.,
\begin{equation}
\frac{\partial}{\partial t}\phi = \frac{3 M}{16 \eta a^3}+\xi(t)\ ,
\end{equation}
where $\phi$ describes the position of the charge relative to the $z$-axis, $M$
is the excerted torque on the particle due to Coulomb forces, and $\xi$ means
the brownian rotational displacement. 

When the system consists of two particles, each carrying one charge
(of opposite sign), that are located at a specified point on the
surface, then the Coulomb force gives a contribution to the rotation
as well as to the translation of the particles. The Coulomb force tends
to rotate the two particles so that the charges approach the common
axis of the particles, the minimum distance.

\begin{figure}
\includegraphics[width=70mm]{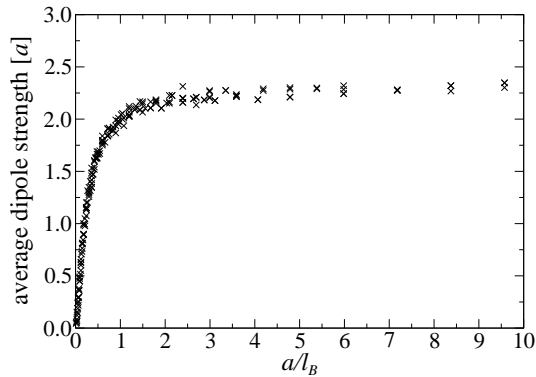}
\caption{
\label{FIGX} \footnotesize
Agglomeration of two spherical particles:
Average dipole strength in units of particle radii $a$ as function of
the particle radius in units of the Bjerrum length $l_B$. For particle
radii bigger than the Bjerrum length, the position of charges during
agglomeration is randomized due to Brownian rotation.
} 
\end{figure}      

Rotational Brownian motion tends to randomize the orientation of the
particles. This means that it is the relative strength of Coulomb force and
Brownian motion that decide whether the two charges find each other upon the
collision, or whether a permanent dipole
pertains. The natural scale for Coulomb energy is the energy of the two
charges two particle radii apart $E_C=q^2/2a$. Rotational Brownian motion
is controlled by thermal energy, and if $k_BT>>E_C$ the
orientation of the particles is completely random when they collide. The
average distance of the resulting dipole can in this case, by a simple
numerical integration over the surfaces of the disks or the spheres, be
calculated. The numerical values are: $2.26 a$ (2D) and $2.33 a$ (3D), i.e.,
slightly more than two particle radii. In the other limit of vanishing
Brownian motion the average distance is of course zero. Further, a crossover
in the average distance of the dipole is expected roughly at $k_BT/E_C=1$.
In colloidal sciences, the crossover between regions dominated by
coulomb interaction on the one hand and by thermal diffusion of
particles on the other hand is often
determined by the Bjerrum length
\begin{equation}
l_B=\frac{q^2}{4\pi\varepsilon\varepsilon_0 k_BT}.
\end{equation}
On distances smaller than the Bjerrum length, interaction of particles
is guided by coulomb interaction, while on larger distances diffusion
dominates. If all colloidal particles carry identical charges, the
suspension is stabilized by Coulomb repulsion if the particle radii
are smaller than the Bjerrum length.

The values of the limits and the location of the cross-over are verified by
numerical experiments as shown in Fig.~\ref{FIGX}. Experiments are performed for
differently sized particles and for different charges in 2D. The data sets
collapse since only the ratio of the Coulomb
energy and the thermal energy is relevant. This universal curve
shows a cross-over, seperating a regime where the residual dipole moment
increases lineary with thermal energy, from a regime where it approaches the
maximal value. This allows one to estimate the typical dipole length, in
experiments as well as in simulations. 

In reality, particles are not perfectly circular or spherical. The roughness of
the surface of a particle has the qualitative implication that rotations
become increasingly difficult with increasing roughness. As a consequence, we
expect a shift of the cross-over to lower thermal energies. 
In this sense, the oberserved point can be considered as an upper
limit for the range where the Coulomb forces dominate.

\section{Conclusions}

The use of electrostatic forces strongly affects the dynamics of
agglomeration processes, for instance coating of micrometer particles
with nanometer particles. In particular, it is still an open question to
what extent the exact locations of charges on the particles have
impact on the resulting agglomerated structures. In this context the
subquestion arises whether recombination of two particles carrying opposite charge
gives a pertaining dipole or whether the charges cancel out. To
simulate large systems of suspended charged particles, knowledge about
charge recombination allows for proper approximations, which in turn
may increase the range of many-particle simulations. 

A detailed study of two idealized situations have been
presented. First the case of one very large particle with a target
charge, which attracts a much smaller particle with one oppsite
charge. Formulating the problem in terms of a Fokker-Planck equation, the
stationary solution is found analytically in 2D and 3D. The
findings are supported by numerical simulations, and the key results
are as follows. Interpreted as a hitting probability distribution,
there is a finite fraction of charged particles that exactly recombine
with the target charge, giving rise to a $\delta$-peak in the distribution. This is valid in 2D
as well as in 3D, although the strength of the $\delta$-peak is somewhat larger
in 2D. However, in 3D there is an additional contribution decaying
like $~1/2r$,
where $r$ is the distance from the target charge. This is the main
difference between 2D and 3D, and it is the reason why superposition
of the solution is possible in 2D, but not in 3D. Further, there is a
constant background term in the probability distribution whose
physical reason lies in the diffusion of the particles. 

Numerics is performed to test the analytic results, and to provide
qualitative insight into the situation of having more target charges
in 3D, which from the point of view of applications is very
important. In this case the superposition of a homogenous background
plus a $1/2r$ shoulder located around each fixed charge fits the
numerical data suprisingly well. Deviations from the superposition 
of single charge solutions are visible in an increasing strength of
delta peaks. Thus, an increased number of fixed charges increases the probability for a
compensation of fixed charges and incoming particles. 

The second idealized situation is the study of two particles of equal
size recombining. In this case the rotational degree of freedom is
included. From physical reasoning and from simulation, we demonstrate
that there is a cross-over when Coulomb energy equal to thermal
energy. Larger thermal energies means a regime where dipoles are
created with random distance. For lower thermal energies the average
length of the dipole goes to zero in the limit, but for finite ratios
of the energies, the average length increases rapidly with the energy
ratio. This implies that for simulations or experiments in this range,
the detailed localization of charges on the particles must be taken
into account for a correct description of the physics.


\vspace{5mm}
\noindent
{\bf Acknowledgements:}\\
We thank S. Dammer, Z. Farkas, M. Linsenb\"uhler, K.-E. Wirth, and D. Wolf 
for fruitful discussions. This work was supported by the German Science
Foundation (DFG) within the research program ''Verhalten Granularer Medien'',
project Hi/744.


\appendix
\section{Solution of the Fokker-Planck equation in two dimensions}

\label{Sol2D}
%
In order to solve the stationary Fokker-Planck equation
\begin{equation}
\nabla^2 P -
\vec{\nabla} P\cdot \uvec-
P \, (\vec{\nabla} \cdot \uvec)=0
\label{EQSTATIONARYFP}
\end{equation}
with $\uvec{=}-\rvec/r^3$, we consider a stationary flow of particles 
inserted at infinity and removed whenever they touch the wall
(cf. Sec.~\ref{SolFokker}). 
Let us first consider the two-dimensional case in the $x,z$-plane,
where the $z$-axis denotes the direction perpendicular to the
surface of the grain (see Fig.~\ref{FIG4}). Introducing
polar coordinates $z{=}r \cos\varphi$ and $x{=}r \sin\varphi$
the gradient acting  on a scalar $f(r,\varphi)$ and the 
divergence of a vector field $\uvec(r,\varphi)$ are
given by
\begin{eqnarray}
\vec{\nabla} f &=& \vec{e}_r\cdot\frac{\partial f}{\partial
  r} +\vec{e}_{\varphi}\cdot \frac{1}{r}\frac{\partial f}{\partial \varphi}\\
\vec{\nabla}\cdot\uvec &=& \frac{\partial }{\partial r} u_r +
\frac{1}{r}\frac{\partial}{\partial \varphi} u_\varphi +
\frac{1}{r} u_r \,,
\end{eqnarray}
where the last term is due to the covariant derivative of vector fields in polar coordinates.
In particular, the Laplacian is given by
\begin{equation}
\nabla^2 = \frac{\partial^2}{\partial
  r^2}+\frac{1}{r^2}\frac{\partial^2}{\partial\varphi^2}+\frac{1}{r}\frac{\partial}{\partial
  r}.
\end{equation}
Inserting these expressions, the stationary Fokker-Planck equation with $\uvec=\rvec/r^3$ reads
\begin{equation}
\left(r^3\frac{\partial^2}{\partial
    r^2}+r\frac{\partial^2}{\partial \varphi^2}+(r^2+
    r)\frac{\partial}{\partial r}-1 \right)P(r,\varphi)=0.
\label{eq_fp_2d}
\end{equation}
Using the Ansatz
\begin{equation}
P(r,\varphi)=Q(\varphi)\cdot R(r)
\end{equation}
one obtains two separate equations 
\begin{eqnarray}
\left(\frac{\partial^2}{\partial \varphi^2}+\cal C\right)Q(\varphi)&=&0
\label{eq_fp_2d_angular}  \\
\left( r^3\frac{\partial^2}{\partial r^2}+(r^2+
  r)\frac{\partial}{\partial r}-(1+{\cal C} r)\right)R(r)&=&0,
\label{eq_fp_2d_radial}
\end{eqnarray}
where $\cal C$ is the common eigenvalue. Symmetry requires that
$Q(\omega)$ is an even function, and the possible solutions of the
angular equation (\ref{eq_fp_2d_angular}) are given by
\begin{equation}
Q(\varphi)=\cos(\omega\varphi) \;\;\mbox{ with }\;\;{\cal C}=\omega^2 \ .
\end{equation}
As the absorbing wall at $z{=}0$ imposes the boundary condition
$Q(-\pi/2)=Q(+\pi/2)=0$ we have
$$ \omega=1,3,5,\ldots \, \, . $$
Inserting ${\cal C}=\omega^{2}$ into equation (\ref{eq_fp_2d_radial})
we find the solutions
\begin{eqnarray}
\omega=1: R(r) & = & Ar+B(r-1)e^{1/r}\nonumber\\
\omega=3: R(r) & = & A(20r^3+8 r^2+  r) \nonumber\\
 & & +B(60r^3-36 r^2+9 r-1)e^{1/r}\nonumber\\
\vdots \;\;\;\;\;\;\;\;\;\;\;\;\;\,& \vdots & \nonumber
\end{eqnarray}
In general these solutions are of the form 
\begin{equation}
A\,f_1(r)+B\,f_2(r)\;e^{1/r} \ ,
\end{equation}
where $f_1$ and $f_2$ are polynomials of degree $\omega$.

Far away from the target charge the probability distribution $P(r,\varphi)$
will not be influenced by the Coulomb force. Since it is
assumed that particles are inserted homogenously at large distance,
we therefore expect a linear asymptotic behavior
\begin{equation}
P(x,z)\propto z \;\;\;\;\mbox{ for }\;\;\;\; z \rightarrow \infty.
\end{equation}
Obviously this condition can only be satisfied for $\omega=1$. 
Furthermore, since $P$ must be positive for small $r$, we find $B{=}0$. Hence
the solution of the two-dimensional Fokker-Planck
equation simply reads
\begin{equation}
P(x,z)=A\cdot z
\end{equation}
independent of the surface charge,
where $A$ is a normalization factor equal to the rate of inserted
particles per unit area at the source plane. Setting $A{=}1$ the
corresponding probability current $\jvec = P \uvec-\vec{\nabla} P
$ for $r{>}0$ is given by 
\begin{equation}
\jvec  = -\frac{xz}{r^3} \;\vec{e}_x
       - \left(1+\frac{z^2}{r^3}\right) \; \vec{e}_z\;.
\label{eq_j_2d}
\end{equation}

\section{Solution of the Fokker-Planck equation in three dimensions}
\label{Sol3D}

Following the previous calculation, we solve the Fokker-Planck equation
(\ref{EQSTATIONARYFP}) by first transforming it to spherical coordinates
\begin{eqnarray}
x & = & r \sin\vartheta\, \cos\varphi \nonumber\\
y & = & r \sin\vartheta\, \sin\varphi \nonumber\\
z & = & r \cos\vartheta\,. \nonumber
\end{eqnarray}
In these coordinates the gradient acting on a scalar is given by
\begin{equation}
\vec{\nabla} f=\vec{e}_r\cdot\frac{\partial f}{\partial
    r}+\vec{e}_{\vartheta}\cdot\frac{1}{r}\frac{\partial f}{\partial \vartheta} 
+\vec{e}_{\varphi}\cdot\frac{1}{r\sin\vartheta}\frac{\partial f}{\partial\varphi}
\end{equation}
while the Laplacian takes the form
\begin{equation}
\nabla^2  =  \frac{\partial^2}{\partial
   r^2}+\frac{2}{r}\frac{\partial}{\partial
   r}+\frac{1}{r^2}\frac{\partial^2}{\partial
   \vartheta^2}+\frac{1}{r^2\tan\vartheta}\frac{\partial}{\partial
   \vartheta}+\frac{1}{r^2\sin^2\vartheta}\frac{\partial^2}{\partial\varphi^2}
\end{equation}
Using again a separation ansatz 
\begin{equation}
P(r,\vartheta,\varphi)=R(r)\cdot
Q(\vartheta,\varphi)\ ,
\end{equation}
we are led to the equations
\begin{eqnarray}
\left(\frac{\partial^2}{\partial
    \vartheta^2}+\frac{1}{\tan
    \vartheta}\frac{\partial}{\partial\vartheta}+\frac{1}{\sin^2
    \vartheta}\frac{\partial^2}{\partial
    \varphi^2}+{\cal C}\right)Q(\vartheta,\varphi)&=&0 \qquad \\
\left( r^2\frac{\partial^2}{\partial
    r^2}+(2r+1)\frac{\partial}{\partial
    r}-{\cal C}\right)R(r)&=&0
\end{eqnarray}
As illustrated in Fig.~\ref{FIG4} the system is invariant
under rotations around the $z$-axis. Thus, the solution will only depend
on $r$ and $\vartheta$, hence $Q(\vartheta,\varphi)=Q(\vartheta)$. 

Solving the angular equation, the general solution
can be expressed in terms of Legendre polynomials. However, for large
$r$ we expect the solution to be independent of the Coulomb field, i.e.,
linear in $z$. Therefore, the only solution of the angular equation,
which satisfies the boundary condition $Q(\pi/2)=0$, turns out to be
$Q(\vartheta)=\cos\vartheta$ with the eigenvalue ${\cal C}=2$.
The corresponding radial equation has the solution
\begin{equation}
R(r)=A (2r+1)+B(2r-1)\,e^{1/r}.
\end{equation}
Since $P(r,\vartheta,\varphi)$ has to be non-negative for small $r$ 
the second term has to vanish, i.e., $B=0$. Choosing $A=1/2$ the
physically meaningful solution reads:
\begin{equation}
P=\left(1+\frac{1}{2r}\right)r\cdot\cos\vartheta=\left(1+\frac{1}{2r}\right)z\ .
\end{equation}
The corresponding probability current $\jvec = P \uvec-\vec{\nabla} P $ 
for $r>0$ is given by
\begin{equation}
\jvec  =   -\Bigl(1+\frac{1}{2r}\Bigr)\evec_z
-\frac{(1+r)}{2}\,\frac{z}{r^4}
\rvec\;.
\label{eq_j_3d}
\end{equation}

\section{Fluxlines of the probability current $\jvec$}
\label{SolFlux}

In two dimensions the trajectories of the vector field~(\ref{eq_j_2d}) can be obtained by solving the differential equation
\begin{equation}
\frac{dz}{dx}=\frac{r^3+z^2}{xz}\,,
\end{equation}
leading to the solution
\begin{equation}
z(x) = \pm\frac{x\sqrt{1-(x+c)^2}}{x+c}\,,
\end{equation}
where $c$ is an integration constant labeling different curves. For the separatrix the slope at the origin
\begin{equation}
\left.\frac{dz(x)}{dx}\right|_{x=0} = \pm\frac{\sqrt{1-c^2}}{c}
\end{equation}
vanishes, i.e. $c=\pm 1$. Selecting the physically meaningful branch the separatrix is given by
\begin{equation}
z(x) = \frac{x \sqrt{x(2-x)}} {1-x}  \qquad (0 \leq x < 1) \,.
\end{equation}
In three dimensions the separatrix can be calculated in the same way. Because of rotational invariance in the $xy$-plane we set $y=0$ so that the trajectories of the vector field~(\ref{eq_j_3d}) obey the differential equation
\begin{equation}
\frac{dz}{dx}=\frac{2r^4+r^3+(1+r)z^2}{(1+r)xz}\ ,
\end{equation}
where $r^2=x^2+z^2$, or equivalently
\begin{equation}
\frac{dz}{dr} = \frac{2r^4+r^3+(1+r)z^2}{rz(2r^2+2r+1)} \,.
\end{equation}
The solution reads
\begin{equation}
z(r) = r \sqrt{\frac{C+2r+2r^2}{1+2r+2r^2}}\ ,
\end{equation}
where $C$ is an integration constant. Since $z'(0){=}0$ implies $C{=}0$ the separatrix in original coordinates is given by
\begin{equation}
z(x) = \frac{x\sqrt{2}}{2x^2-1} \, \sqrt{2(x^2-x^4)+\sqrt{x^2-x^4}}\ .
\end{equation}

\section{Particle density at the wall}
\label{SolWall}

The main quantity of interest, which can be calculated from the
probability density flux, is the distribution of the particles that
hit the wall. Due to the rotational symmetry of the configuration
around the charged particle, this density, $\rho(r)$, is a function of
the radial distance $r$ from the charge only. The density is equal to
the normal component of the flux at the wall 
\begin{equation}
  \rho(r) = \left. -j_z(\rvec) \right|_{z=0}\ .
\end{equation}
For all points on the wall, except for $r{=}0$, it follows by direct
insertion into Eqs. (\ref{eq_j_2d}) and (\ref{eq_j_3d}), that 
\begin{equation}
\rho(r) =
\left\{
\begin{array}{ll}
1        & \text{in 2D} \\
1 + 1/2r & \text{in 3D}
\end{array} \qquad (r > 0)\ .
\right.
\end{equation}

A certain fraction of the particles hit the target charge directly at
$r{=}0$, thus giving a $\delta$-peak contribution at this point. In 2D
the strength of the peak is 
\begin{eqnarray}
  \lim\limits_{R\to 0} \left( -\int\limits_{-\pi/2}^{\pi/2}
  {\hat{n}\cdot\jvec\,R\,d\varphi} \right) &=&
  \lim\limits_{R\to 0} \left( -\int\limits_{-\pi/2}^{\pi/2}
  \cos{\varphi} \left[R+1\right]\,d\varphi \right)
  \nonumber \\
  &=& \lim\limits_{R\to 0} 2(R+1) = 2\ ,
\end{eqnarray}
where $\hat{n}{=}\vec{e}_z\cdot\cos{\varphi}+\vec{e}_x\cdot\sin{\varphi}$ is
the normal vector on a half-sphere over which the integral in-flux of
particles is calculated. The actual in-flux at $r{=}0$ is found in the
limit $r{\to}0$.

Similarily in 3D, by taking the unit normal vector as
$\hat{n}=(\sin\vartheta\cos\varphi,\sin\vartheta\sin\varphi,\cos\varphi)$,
one obtains 
\begin{eqnarray}
  & & \lim\limits_{R\to 0} \left( - \int\limits_0^{\pi/2}d\vartheta\int\limits_0^{2\pi}
  d\varphi R^2\sin\vartheta\,\hat{n}\cdot\jvec \right) \nonumber \\
  & = & \lim\limits_{R\to 0}
  -2\pi\left(R^2+R+\frac{1}{2} \right)
  \int\limits_{0}^{\pi/2}\cos\vartheta\sin
  \vartheta\,d\vartheta  \nonumber\\
 & = & \frac{\pi}{2}   \ ,
\end{eqnarray}
where the integration is taken over a half-sphere around the origin with radius $R$.
Combining these results we arrive at 
\begin{equation}
\rho(r) =
\left\{
\begin{array}{ll}
1 + 2\delta(r)  & \text{in 2D} \\
1 + 1/2r + \frac{\pi}{2}\delta(r) & \text{in 3D}\ .
\end{array}
\right.
\end{equation}
In the 2D solution it is understood that the $\delta$-function
integrated over the target line gives unity. Likewise in 3D,
integration of the $\delta$-function over the target plane gives unity.


\end{document}